\documentclass{JHEP3} 

\usepackage{epsfig,multicol}
\usepackage{amsmath, amssymb}
\usepackage{concmath, palatino}

\newcommand\fverb{\setbox\pippobox=\hbox\bgroup\verb}
\newcommand\fverbdo{\egroup\medskip\noindent%
			\fbox{\unhbox\pippobox}\ }
\newcommand\fverbit{\egroup\item[\fbox{\unhbox\pippobox}]}
\newbox\pippobox

\newcommand{\be}{\begin{equation}}
\newcommand{\ee}{\end{equation}}
\newcommand{\ba}{\begin{eqnarray}}
\newcommand{\ea}{\end{eqnarray}}

\def\M#1{{{\rm\bf M}}\left[#1\right]}
\def\vare{\varepsilon}
\def\bi{\bibitem}

\title{Four loop reciprocity of twist two operators \\ in $\mathcal{N}=4$ SYM}

\author{Matteo Beccaria\\
  Physics Department, Salento University, 
  Via Arnesano, 73100 Lecce\\
  INFN, Sezione di Lecce\\
  E-mail: \email{matteo.beccaria@le.infn.it}}

\author{Valentina Forini\\
  Max-Planck-Institut f\"ur Gravitationsphysik,
  Albert-Einstein-Institut, \\
  Am M\"uhlenberg 1, D-14476 Potsdam, Germany \\
  E-mail: \email{forini@aei.mpg.de}}

\preprint{AEI-2009-001}

\abstract{
The four loop universal anomalous dimension of twist-2 operators in $\mathcal{N}=4$ SYM has been recently conjectured. 
In this paper, we prove that it obeys a generalized Gribov-Lipatov reciprocity, previously known to hold at the three loop level. 
}

\begin{document} 

\section{Introduction}

The four loop universal anomalous dimension of $\mathcal{N}=4$ SYM twist $\tau=2$ operators with general spin  has been conjectured in~\cite{KLRSV,Janik}. 
In perturbation theory, the spin $N$ dependent anomalous dimension $\gamma(N)$ is the sum of two pieces, called the {\em asymptotic} and {\em wrapping} contributions.
The asymptotic term can be computed rigorously for each $N$ by means of the  all loop asymptotic Bethe Ansatz of $\mathcal{N}=4$ SYM~\cite{Beisert:2005fw}. 
The wrapping correction  starts at loop order $\tau+2$ and is currently believed to be correctly predicted by generalized L\"{u}scher formulas~\cite{Janikprev}.

At three loops, the sequence of values $\{\gamma(N)\}_{N=1, 2, \dots}$ can be expressed as a closed function of the spin $N$~\cite{KLOV,factorized,Kotikov:2008pv} as a linear
combination of nested harmonic sums.
 This exact result is in agreement with the QCD-inspired maximum transcendentality principle~\cite{MTP} which is usually accepted to hold at any loop order. This leads to the four loop conjecture of~\cite{KLRSV,Janik} which, as a check, reproduces the correct cusp anomaly
as well as the pole structure predicted by the next-to-leading order BFKL equations~\cite{BFKL}.

Another property,  known to be valid at three loops,  emerges as a higher order generalization of the one-loop Gribov-Lipatov reciprocity~\cite{dok1,bk,dok2}.
 The crossed QCD processes of deep inelastic scattering  and $e^+e^-$ annihilation can be treated symmetrically in an approach 
based on modified DGLAP evolution equations for parton distributions with kernel $P(N)$ obeying perturbatively
\be
\label{nonlinear}
\gamma(N) = P\left(N+\textstyle{\frac{1}{2}\gamma(N)}\right).
\ee
The reciprocity condition is a constraint on  the large spin $N$ behavior of $\gamma(N)$~\cite{bk} which can be written 
as the following asymptotic condition on $P(N)$
\be
\label{RRJ}
P(N) = \sum_{\ell\ge 0} \frac{a_\ell(\log\,J^2)}{J^{2\,\ell}}, \quad J^2 = N\,(N+1),
\ee  
where $a_\ell$ are suitable {\em coupling-dependent} polynomials. 
$J^2$ is the Casimir of the collinear subgroup $SL(2, \mathbb{R})\subset SO(2,4)$ of the  conformal  group~\cite{bra}
and the above constraint is simply parity invariance under (large) $J\to -J$.
A generic expansion around $N=\infty$ can involve odd powers of $1/J$. These are  forbidden in Eq.~(\ref{RRJ}).

We remark that reciprocity is not a rigorous prediction. Instead, it is a property which is based on sound physical arguments and deserve to be 
tested at higher loop order. Its persistent validity is an intriguing empirical observation which deserves a deeper understanding. Indeed it has been observed in several 
QCD and $\mathcal{N}=4$ SYM multi-loop calculations~\cite{Forini:2008ky}. In particular, three-loop tests of reciprocity for QCD and for the universal
 twist 2 supermultiplet in $\mathcal{N}=4$ SYM were discussed in~\cite{bk,dok2}, a four-loop test for
the   twist 3  anomalous dimension was performed  in~\cite{bdm} in the scalar sector, and in~\cite{bf} in the "gluon" sector.
 All these examples did not require the wrapping correction.

Here, we show that the twist-2 four loop result is again reciprocity respecting, including wrapping, although certain additional {\em simplicity} features 
of the three loop result will be shown to be broken.  Finally, we refer the reader to the five loop, twist three results of  Ref.~\cite{BFLZ}, which show strict analogies with the present case for the presence of a wrapping contribution,  the reciprocity analysis and the asymptotic features.

\section{Proof of reciprocity}
\label{sec:proof}

The twist-2 anomalous dimension is  written perturbatively as
\be
\gamma(N) = g^2\,\gamma_1(N) + g^4\,\gamma_2(N) + g^6\,\gamma_3(N) +g^8\,\big( \gamma_4^{ABA}(N) + \gamma_4^{\rm wrapping}(N)\big) + {\cal O}(g^{10}),
\ee
where $g^2=\frac{g^2_{\rm  YM} N}{16\,\pi^2}$. For $\gamma_1(N),\,\gamma_2(N),\,\gamma_3(N)$, explicit formulas in terms of linear combinations of harmonic sums  can be found in Refs.~\cite{KLOV,factorized,Kotikov:2008pv}. $\gamma_4^{ABA}(N)$ is the asymptotic Bethe Ansatz result reported in Table 1 of ~\cite{KLRSV}, while  the wrapping contribution $ \gamma_4^{\rm wrapping}(N)$
can be found in~\cite{Janik}.
 
The $P$-kernel defined via (\ref{nonlinear}) can be derived from the anomalous dimension by simply inverting (\ref{nonlinear}). 
Expanding perturbatively $P$ as
\be
P(N) = g^2\,P_1(N) +g^4\,P_2(N) +g^6\,P_3(N) +g^8\,P_4(N) + {\cal O}(g^{10}),
\ee
one finds the four loop contribution to be
\ba\label{P4}
P_4 &=& P_4^{ABA} + P_4^{\rm wrapping}, \\
P_4^{ABA} &=& \gamma _4^{ABA} -\frac{1}{48} \gamma _1^{(3)} \gamma _1^3-\frac{3}{16} \gamma _1' \gamma _1'' \gamma _1^2+\frac{1}{8} \gamma _2'' \gamma _1^2-\frac{1}{8} \left(\gamma _1'\right)^3
   \gamma _1+\frac{1}{2} \gamma _1' \gamma _2' \gamma _1-\frac{1}{2} \gamma _3' \gamma _1+\nonumber\\
&& +\frac{1}{4} \gamma _2 \gamma _1'' \gamma _1+\frac{1}{4} \gamma _2
   \left(\gamma _1'\right)^2-\frac{1}{2} \gamma _3 \gamma _1'-\frac{1}{2} \gamma _2 \gamma _2', \\
P_4^{\rm wrapping} &=& \gamma_4^{\rm wrapping}.
\ea
The above expression is an explicit linear combination of products of harmonic sums, that for the purpose of proving reciprocity it is  useful to rewrite  in a canonical basis, \emph{i.e.} as linear combinations of single sums. This can be done by repeatedly using the shuffle algebra relation (\ref{eq:shuffle}).

The proof that $P_4$ is reciprocity respecting  is based on rewriting Eq.(\ref{P4})  in terms of special linear combinations of harmonic sums with definite properties under the (large-)$J$ parity $J\to -J$. We introduce them in the following section.

\subsection{Definite-parity linear combinations of harmonic sums}
\label{sec:omega}

Let us consider the space $\Lambda$ of $\mathbb{R}$-linear combinations of harmonic sums $S_\mathbf{a}$ with generic
multi-indices 
\be
\mathbf{a} = (a_1, \dots, a_\ell),\qquad a_i\in \mathbb{Z}\backslash \{0\},
\ee
where $\ell$ is not fixed. At any perturbative order, $P\in \Lambda$.

For any $a\in\mathbb{Z} \backslash  \{0\}$, we define the linear map $\omega_a:\Lambda\to\Lambda$ by assigning 
its action on single harmonic sums
as follows
\be\label{map}
\omega_a(S_{b, \mathbf{c}}) = S_{a, b, \mathbf{c}}-\frac{1}{2}\,S_{a\wedge b, \mathbf{c}},
\ee
where, for $n, m\in \mathbb{Z} \backslash  \{0\}$, the wedge-product is
\be\label{wedge}
n\wedge m = \mbox{sign}(n)\,\mbox{sign}(m)\,(|n| + |m|).
\ee
Besides basic harmonic sums, it is  convenient to work with the complementary sums $\underline{S_\mathbf{a}}$ defined in Appendix ~A. On the space $\underline{\Lambda}$ 
of their $\mathbb{R}$-linear combinations  a linear map $\underline{\omega_a}$ can be defined in total analogy with (\ref{map}).

\noindent In the spirit of~\cite{dok2,bdm}, we  introduce the combinations of (complementary) harmonic sums
\be
\begin{array}{lll}
\Omega_a & = & S_a, \\
\Omega_{a, \mathbf{b}} & = & \omega_a( \Omega_\mathbf{b}),
\end{array}
\qquad
\begin{array}{lll}
\underline{\Omega_a} & = & S_a = \underline{S_a}, \\
\underline{\Omega_{a, \mathbf{b}}} & = & \underline{\omega_a}( \underline{\Omega_\mathbf{b}}).
\end{array}
\ee
for which the following two theorems,  proved  in App.~\ref{sec:thm}, hold.

\bigskip
\noindent
{\bf Theorem 1:} ~\footnote{A special case of Theorem 1 appeared in~\cite{dok2}. A general proof of Theorem 1 in the restricted case 
$\mathbf{a} = (a_1, \dots, a_\ell)$ with {\em positive} $a_i>0$ and {\em rightmost indices} $a_\ell\neq 1$ can be found in~\cite{bdm}. 
Appendix~\ref{sec:thm} contains the proof of the general case.} {\em The subtracted complementary  combination $\underline{\widehat\Omega_{\mathbf{a}}}$, $\mathbf{a} = (a_1, \dots, a_d)$ has definite parity 
$\cal P_\mathbf{a}$
under the (large-)$J$ transformation  $J\to -J$ and
\be
{\cal P}_\mathbf{a} = (-1)^{|a_1|+\cdots + |a_d|}\,(-1)^d\,\prod_{i=1}^d \varepsilon_{a_i}.
\ee
}
\noindent
{\bf Theorem 2:} {\em The combination $\Omega_{\mathbf{a}}$, $\mathbf{a} = (a_1, \dots, a_d)$ with odd positive $a_i$ and 
even negative $a_i$ has positive parity ${\cal P}=1$.}
 
Theorem 2 follows from Theorem 1 (see Appendix ~B). In this paper we shall use the second theorem  only, but we quote
the first as a separate result since it can be relevant in more involved situations~\cite{BFLZ}.

\subsection{The four loop $P$-kernel}

The strategy to prove  reciprocity for the kernel  $P$ is the following.  For each loop order $\ell$, one considers in $P_\ell$ (written in the canonical basis) the sums  with maximum depth, each of them, say $S_\mathbf{a}$, appearing uniquely as the maximum depth term in $\Omega_\mathbf{a}$. One then subtracts all the $\Omega$'s required to cancel these terms, keeps track of this subtraction and repeats the procedure with depth decreased by one.
At the end, if the remainder is zero and if the full subtraction is composed of $\Omega$'s with the right parities, as prescribed by Theorem 2, 
we have proved that $P$  is reciprocity respecting.
This reduction algorithm can be successfully applied up to three loops, here we report the four loop case.

The expression for $P_4^{ABA}$ in the canonical basis is very long, and we do not show it.
Applying the reduction algorithm one find the following form of the asymptotic contribution
\ba\label{P4ABA}
P_4^{ABA} &=& -8192 \,\Omega _{1, 1, 1, -2, 1, 1}
+ 6144 \,\Omega _{-2, -2, 1, 1, 1} + 6144 \,\Omega _{-2, 1, -2, 1, 1} + 
  4096 \,\Omega _{-2, 1, 1, -2, 1} + \nonumber\\
  &&+ 6144 \,\Omega _{1, -2, -2, 1, 1} 
  6144 \,\Omega _{1, -2, 1, -2, 1} + 2048 \,\Omega _{1, -2, 1, 1, -2} + 
  6144 \,\Omega _{1, 1, -2, -2, 1} + \nonumber\\
&& +4096 \,\Omega _{1, 1, -2, 1, -2} + 
  6144 \,\Omega _{1, 1, 1, -2, -2}
-1024 \,\Omega _{-2, -2, -2, 1} - 1536 \,\Omega _{-2, -2, 1, -2} +\nonumber\\
&& - 
  2048 \,\Omega _{-2, 1, -2, -2} + 1024 \,\Omega _{1, -4, 1, 1} - 
  1536 \,\Omega _{1, -2, -2, -2} + 3072 \,\Omega _{1, 1, -4, 1} + \nonumber\\
  &&+ 1024 \,\Omega _{1, 1, -2, 3} +2048 \,\Omega _{1, 1, 1, -4} + 
  2048 \,\Omega _{1, 3, -2, 1} + 1024 \,\Omega _{3, -2, 1, 1} + 
  2048 \,\Omega _{3, 1, -2, 1} + \nonumber\\
&&
-2048 \,\Omega _{-4, -2, 1} - 1280 \,\Omega _{-4, 1, -2} - 
  2048 \,\Omega _{-2, -4, 1} - 768 \,\Omega _{-2, -2, 3} - 
  1536 \,\Omega _{-2, 1, -4} +\nonumber\\
&& - 256 \,\Omega _{-2, 3, -2} - 
  2304 \,\Omega _{1, -4, -2} - 1792 \,\Omega _{1, -2, -4} - 
  2048 \,\Omega _{1, 1, 5} - 1536 \,\Omega _{1, 5, 1} +\nonumber\\
&& - 
  1280 \,\Omega _{3, -2, -2} - 1536 \,\Omega _{5, 1, 1}
-768 \,\Omega _{-6, 1} - 128 \,\Omega _{-4, 3} + 384 \,\Omega _{-2, 5} - 
  1408 \,\Omega _{1, -6} +\nonumber\\
&& - 896 \,\Omega _{3, -4} - 256 \,\Omega _{5, -2}
+640 \,\Omega _7
+\frac{2048}{3} \pi ^2 \,\Omega _{1, 1, -2, 1} + 
  1024 \pi ^2 \,\Omega _{1, 1, 1, -2} + \nonumber\\
&&
-\frac{512}{3} \pi ^2 \,\Omega _{-2, -2, 1} - \frac{512}{3} \pi ^2 \,\Omega _{-2, 
      1, -2} - \frac{512}{3} \pi ^2 \,\Omega _{1, -2, -2}
-\frac{256}{3} \pi ^2 \,\Omega _{-4, 1} +\nonumber\\
&& - 
  256 \pi ^2 \,\Omega _{1, -4} - \frac{512}{3} \pi ^2 \,\Omega _{3, -2}
+1536 \zeta_3 \,\Omega _{-2, 1, 1} + 1280 \,\Omega _{1, -2, 1} \zeta_3 + 
  1024 \,\Omega _{1, 1, -2} \zeta_3 \nonumber\\
&& 
+640 \zeta_3 \,\Omega _{1, 3} + 640 \,\Omega _{3, 1} \zeta_3
-320 \,\Omega _{-4} \zeta_3 
+\frac{1088}{15} \pi ^4 \,\Omega _{1, 1, 1}
-\frac{64}{3} \pi ^4 \,\Omega _{1, -2}
-\frac{752}{45} \pi ^4 \,\Omega _3 + \nonumber\\
&& 
+\,\Omega _{1, 1} \big(-\frac{256}{3} \pi ^2 \zeta_3 + 2560 \zeta_5\big)
-\frac{256}{45} \pi ^6 \,\Omega _1-\zeta_3(  2\,\Omega_{-2,1} + \Omega_3),\\\nonumber
\ea
while the wrapping contribution reads
\be\label{P4wr}
P_4^{\rm wrapping} = -128\,\Omega_1^2\,(5\,\zeta_5+4\,\zeta_3\,\Omega_{-2}+8\,\Omega_{-2,-2,1}+4\,\Omega_{3, -2}).
\ee
This proves reciprocity at four loops, since in the expressions above only allowed $\Omega$'s appear. Notice that the asymptotic and wrapping contributions are \emph{separately} reciprocity respecting, an interesting feature which happens to be present also in the twist three five-loop analysis of~\cite{BFLZ}.

\section{Expansions at large $N$ and inheritance violation}
\label{sec:expansions}

The general structure of soft gluon emission governing the very large $N$ behaviour of $\gamma(N)$ predicts the leading contribution $\gamma(N)\sim f (\lambda)\,\log\,N$
where the coupling dependent scaling function $f (\lambda)$ (cusp anomaly) is expected to be universal in {\em both twist and flavour}~\cite{Belitsky:2003ys,Eden:2006rx} .
This is precisely what is observed in the various exact multiloop expressions discussed in Appendix~F of~\cite{BFTT}.

This leading logarithmic behaviour is also the leading term in the function $P(N)$. Concerning the subleading terms, 
as remarked in ~\cite{bk,dok2},  the function $P(N)$ obeys at three loops 
a  powerful additional {\em simplicity} constraint, in that it does not contain  logarithmically enhanced 
terms $\sim \log^n(N)/N^m$ with $n\ge m$. This immediately implies that the leading logarithmic functional relation 
\be\label{resum}
 \gamma(N)=f (\lambda)\,\log\left(\textstyle{N+\frac{1}{2}f (\lambda)\log N+} ...\right)+...
 \ee
predicts correctly the maximal logarithmic terms $\log^m{N}/N^m$ 
\be\label{leadinglogs}
\gamma(N)\sim f\,\log N+\frac{f ^2}{2}\,\frac{\log N}{N}-\frac{f ^3}{8}\,\frac{\ln^2 N}{N^2}+...
\ee
whose coefficients are simply proportional to $f ^{m+1}$~\cite{bdm,bkp,BFTT}.

Notice that the fact that the cusp anomaly is  known at all orders in the coupling via the results of~\cite{bes,bkk} naturally implies 
(\emph{under the ``simplicity'' assumption for  $P$}) a proper \emph{prediction} for such maximal logarithmic terms at \emph{all orders} in the coupling constant, and in particular for those appearing in the large spin 
expansion of the energies of certain semiclassical string configurations (dual to the operators of interest). 
Such prediction has been  checked in~\cite{BFTT}  up to one loop in the sigma model semiclassical expansion, as well as 
in~\cite{Ishizeki:2008tx} at the classical level. An independent strong coupling confirmation 
of (\ref{leadinglogs}) up to order $1/M$   has recently been given for twist-two operators in~\cite{Freyhult:2009my}. 

However, the asymptotic part of the four loop anomalous dimension for twist-2 operators reveals an exception to this "rule", being the term $\log^2 N/N^2$  not given only in terms of the cusp anomaly.
Interestingly enough, the large spin expansion of the wrapping contribution of~\cite{Janik}, which correctly does not change the leading asymptotic behavior (cusp anomaly), 
first contributes at the same order $\log^2 N/N^2$. 
Thus, while on the basis of (\ref{leadinglogs}) one would expect in the large spin expansion of the four loop anomalous dimension a term of the type  (we denote by $(\cdots)_4$ the
4-th loop contribution)
\be\label{predc22}
c\, \frac{\log^2N}{N^2}~~~~~~~~{\rm with}~~~~~~~~(c)_4=\big(\textstyle{-\frac{f^3}{8}}\big)_4=64\,\pi^2
\ee
expanding $\gamma_4$ according to App.~(\ref{app:asym}), one finds instead  (see Appendix~C, formulas (\ref{exp4ABA}) and (\ref{exp4wr}))
\be\label{effc22}
(c^{\rm ABA})_4 = 64\pi^2-128\,\zeta_3~~~~~~~~{\rm and}~~~~~~~~
(c^{\rm wrapping})_4 = -\frac{64}{3}\pi^2-128\,\zeta_3
\ee
which summed up do not reproduce (\ref{predc22}). 
This indicates that, in the case of the twist-2 operators and starting at four loops, the $P$-function ceases to be "simple" in the meaning of ~\cite{dok2}. 
This is confirmed by explicitely looking at the the structure of its asymptotic expansion (formula \ref{P4exp} below), and prevents the tower of subleading logarithmic singularities $\log^m N/N^m$ to be simply inherited from the cusp anomaly. In order to clarify how the observed difference in the simplicity of the $P$ at weak and strong coupling works, further orders  in the semiclassical sigma model expansion would be needed.

\section{Conclusions}

The present analysis together with the related work in~~\cite{dok1,bk,dok2,bdm,Beccaria:2007cn,Beccaria:2007pb,Beccaria:2007vh,bf, BFTT}
leads to the following conclusions.

At weak coupling, reciprocity has been tested at higher loop order in $\mathcal{N}=4$ SYM at weak coupling for the 
minimal dimension of operators of twist $\tau=2$ and $\tau=3$ for all possible flavors.
The present 
paper shows for the first time that this holds true even at wrapping order (see also~\cite{BFLZ}).

At strong coupling, reciprocity can be investigated by employing AdS/CFT correspondence, 
which indicates the folded string as the configuration  dual to twist-2 operators~\cite{folded}.
This analysis, initiated in~\cite{bk} for the folded string at the classical level, has been recently extended in~\cite{BFTT} at one loop in string 
perturbation theory.
Remarkably, the large spin expansion of the string energy does respect reciprocity,
providing a strong indication that these relations hold not only in weak coupling (gauge theory) but also in 
strong coupling (string theory) perturbative expansions.

All this suggests reciprocity to be an underlying property of $\mathcal{N}=4$ SYM. While it would be significative to \emph{derive} it from first principles 
 (it is expected that the AdS/CFT correspondence might help in this),  
a reasonable attitude can be  to just  \emph{assume} it
as a heuristic guiding principle useful to formulate conjectures on closed formulae~\footnote{For example, the conjecture of~\cite{Veliz} for the  coefficient of $\zeta_5$ in the twist-2 anomalous 
dimension at four loops  contains in principle an arbitrary rational number. A precise value was suggested in~\cite{Veliz} on the basis of some 
deep physical intuition and then confirmed in ~\cite{Janik}. That same value would be unambiguously  selected by requiring the reciprocity of the conjecture.}.
The use of both the maximum transcendentality principle \emph{and } reciprocity drastically reduces the number of terms that have to be 
calculated via Bethe Ansatz and generalised L\"{u}scher techniques. A recent example which illustrates this approach is the five-loop anomalous dimension of twist-3 operators 
described in~\cite{BFLZ}.

\section*{Acknowledgments}
We thank N. Beisert, G. Korchemsky, A. Rej, M. Staudacher  and A. Tseytlin for discussions.
M. B. is greatly indebted to Y. Dokshitzer and G. Marchesini for sharing deep ideas on the issues 
of reciprocity.

\appendix
\section{Harmonic sums}
\label{sec:harmonic}

\subsection{Basic definitions}

The basic definition of nested harmonic sums  $S_{a_1, \dots, a_\ell}$ is recursive
\be
S_a(N) = \sum_{n=1}^N\frac{\varepsilon_a^n}{n^{|a|}}, ~~~~~~~~~~~
S_{a, \mathbf{b}}(N) = \sum_{n=1}^N\frac{\varepsilon_a^n}{n^{|a|}}\, S_{\mathbf b}(n),
\ee
where $\varepsilon_a = +1(-1)$ if $a\ge 0$ ($a<0$).
The \emph{depth}  of a given sum $S_\mathbf{a} = S_{a_1, \dots, a_\ell}$ is defined by the integer $\ell$, while its \emph{transcendentality}  is the sum  $|\mathbf{a}|= |a_1| + \cdots + |a_n|$.
Product of $S$ sums can be reduced to linear combinations of single sums by using iteratively the shuffle algebra~\cite{Blumlein} defined as follows
\ba
\label{eq:shuffle}
&& S_{a_1, \dots, a_\ell}(N)\,S_{b_1, \dots, b_k}(N) =\sum_{p=1}^N \frac{\varepsilon_{a_1}^p}{p^{|a_1|}}\,S_{a_2, \dots, a_\ell}(p)\,S_{b_1, \dots, b_k}(p) + \\
&& + \sum_{p=1}^N \frac{\varepsilon_{b_1}^p}{p^{|b_1|}}\,S_{a_1, \dots, a_\ell}(p)\,S_{b_2, \dots, b_k}(p) - \sum_{p=1}^N \frac{\varepsilon_{a_1}^p\,\varepsilon_{b_1}^p}{p^{|a_1|+|b_1|}}\,S_{a_2, \dots, a_\ell}(p)\,S_{b_2, \dots, b_k}(p). \nonumber
\ea

\subsection{Complementary and subtracted sums}

Let $\mathbf{a} = (a_1, \dots, a_\ell)$ be a multi-index. For $a_1\neq 1$, it is convenient to adopt the concise notation
\be
S_\mathbf{a}(\infty)\equiv S^*_\mathbf{a}.
\ee
Complementary harmonic sums are defined recursively by $\underline{S_a} = S_a$ and 
\be
\underline{S_\mathbf{a}} = S_\mathbf{a}-\sum_{k=1}^{\ell-1} S_{a_1,\dots, a_k}\,\underline{S_{a_{k+1},\dots, a_\ell}^*}.
\ee
The definition is ill when $\mathbf{a}$ has some rightmost 1 indices. In this case, we treat $S_1^*$ as a formal object in the 
above definition and set it to zero in the end. Since $\underline{S_\mathbf{a}}^* <\infty$ in all remeining cases, it is meaningful to  define subtracted complementary sums as
\ba
\underline{\widehat{S}_\mathbf{a}} &=& \underline{S_\mathbf{a}} -\underline{S_\mathbf{a}^*}.
\ea
Explicitely,
\be\label{subtracted}
{\underline{\widehat S_{\mathbf{a}}}}(N)=(-1)^\ell\,\sum_{n_1=N+1}^\infty\frac{\vare_{a_1}^{n_1}}{n_1^{|a_1|}}\,\sum_{n_2=n_1+1}^\infty\frac{\vare_{a_2}^{n_2}}{n_2^{|a_2|}}\,\dots\,\sum_{n_\ell=n_{\ell-1}+1}^\infty\frac{\vare_{a_\ell}^{n_\ell}}{n_\ell^{|a_\ell|}}.
\ee

\subsection{Asymptotic expansions of harmonic sums}
\label{app:asym}

We briefly illustrate how to derive the large $N$ expansion of a nested harmonic sums, considering here for simplicity the case of positive indices  (generalization is straightforward).
We first define
\be
S_a^{(p)}(N) = \sum_{n=1}^N\frac{\log^p n}{n^a}.
\ee
In the case of one index, the following expansions hold ($B_k$ are Bernoulli's numbers)
\ba
S_1(N) &=& \log\,N+\gamma_E+\frac{1}{2\,N}-\sum_{k\ge 1}\frac{B_{2\,k}}{2\,k\,N^{2\,k}}, \\
S_a(N) &=& \zeta_a + \frac{a-2\,N-1}{2\,(a-1)\,N^a}-\frac{1}{(a-1)!}\sum_{k\ge 1}\frac{(2\,k+a-2)!\,B_{2\,k}}{(2\,k)!\,N^{2\,k+a-1}}, \qquad a\in \mathbb{N}, a>1.
\ea
The multiple  sums $S_\mathbf{a}$ can be treated as follows. Let $\mathbf{a} = (a_1, a_2, \dots, a_k)$. Suppose that the expansion of 
$S_{a_2, \dots, a_k}(N)$ is known. Its general form will be of the type
\be
S_{a_2, \dots, a_k} = \sum_{p,q} c_{p,q}\, \frac{\log^p N}{N^q},
\ee
and thus
\be
S_\mathbf{a} = \sum_{n=1}^N \frac{1}{n^{a_1}}\,S_{a_2,\dots, a_k}(n) = 
\sum_{p,q} c_{p,q} \sum_{n=1}^N \frac{\log^p n}{n^{a_1+q}}.
\ee
This determines the expansion of the sum apart from the constant term $S_\mathbf{a}(\infty)$, which can be evaluated using  the package in~\cite{HPL}.

\subsection{Mellin transforms}
\label{subsec:mellin}

Let $\mathbf{a}=\{a_1,\dots,a_\ell\}$ be a multi-index with no rightmost indices equal to $1$, $a_\ell\neq 1$. Defining recursively the functions $G(x)$ via
\ba
G_{a_1,\dots,a_\ell}(x)&=&\frac{1}{\Gamma(|a_1|)}\int_x^1\frac{dy}{y-\vare_{a_2}\dots\vare_{a_\ell}}\,\ln^{|a_1|-1}\frac{y}{x}
~~G_{a_2,\dots,a_\ell}(y)\\\nonumber
\dots&~~~&\dots\\\nonumber
G_{a_{\ell-1},a_\ell}(v)&=&\frac{1}{\Gamma(|a_{\ell-1}|)}\int_v^1\frac{dw}{w-\vare_{a_{\ell}}}\,\ln^{|a_{\ell-1}|-1}\frac{w}{v}
~~G_{a_\ell}(w)\\
G_{a_\ell}(w)&=&\frac{1}{\Gamma(|a_\ell|)}\ln^{|a_\ell|-1}\frac{1}{w}
\ea
the Mellin transform of the subtracted sums of (\ref{subtracted}) is then defined via
\ba\nonumber
{\underline{\widehat S_{\mathbf{a} }}}(N)&=&(\vare_{a_1}\dots\vare_{a_\ell})^N\int_0^1dx\,x^{N-1}\, \frac{x}{x-\vare_{a_1}\dots\vare_{a_\ell}}~~G_{a_1,\dots,a_\ell}(x)  \\\label{Mellinsub}
&\equiv& (\vare_{a_1}\dots\vare_{a_\ell})^N\,\M{\frac{x}{x-\vare_{a_1}\dots\vare_{a_\ell}}~~G_{a_1,\dots,a_\ell}(x)}
\ea

For example, for three indices it is 
\be
{\underline{\widehat S_{a,b,c}}}(N)=\frac{(\vare_a\vare_b\vare_c)^{N}}{\Gamma(|a|)\Gamma(|b|)\Gamma(|c|)}\,\,\M{\frac{x}{x-\vare_a\vare_b\vare_c}\int_x^1\frac{dy}{y-\vare_b\vare_c}\ln^{|a|-1}\frac{y}{x}\int_y^1\frac{dz}{z-\vare_c}\ln^{|b|-1}\frac{z}{y}\,\ln^{|c|-1}\frac{1}{z}}
\ee

For our purpose, it is important to notice that the function $G$ in (\ref{Mellinsub})  satisfies the property 
\ba\label{Ggeneral}
&&G_{a_1,\dots,a_\ell}\left(\textstyle{\frac{1}{x}}\right)=\textstyle{(-1)^{\sum_{i=1}^\ell (|a_i|-1)}}\Big\{
G_{a_1,\dots,a_\ell}(x)-
\sum_{k=1}^{\ell-1}\textstyle{G_{a_1,\dots,a_{k}\wedge a_{k+1},\dots,a_\ell}}(x)\\\nonumber
&& 
+\Big[\sum_{k=1}^{\ell-1}G_{a_1,\dots,a_{k-1}\wedge a_{k}\wedge a_{k+1},\dots,a_\ell}(x)
 +\sum_{k=1}^{\ell-2}\textstyle{G_{a_1,\dots,a_{k-1}\wedge a_{k},a_{k+1}\wedge a_{k+2},\dots,a_\ell}}(x)\Big]\\\nonumber
&& -\Big[\,\sum_{k=1}^{\ell-1}\textstyle{G_{a_1,\dots,a_{k-2}\wedge a_{k-1}\wedge a_{k}\wedge a_{k+1}, a_{k+2},\dots,a_\ell}}(x)+\dots\Big]+\dots+(-1)^{\ell -1}\textstyle{G_{a_1\wedge a_2\wedge \dots\wedge a_\ell}}(x)
\Big\}
\ea
Above, the sign of each contribution is determined by $(-1)^{n_{\rm w}}$, with $n_{\rm w}$ is the number of the wedge-products in the $G$-functions appearing in that piece.
For example, for three indices it is
\be
G_{a,b,c}\left(\textstyle{\frac{1}{x}}\right)=(-1)^{|a|+|b|+|c|-1}\,\left[G_{a,b,c}(x)-\,G_{a\wedge b,c}(x)-\,G_{a,b\wedge c}(x)+\,G_{a\wedge b\wedge c}(x)\right]
\ee
To obtain (\ref{Ggeneral}), one uses recursively the result
\be
\frac{1}{\Gamma(|a_1|)}\int_x^1\frac{dy}{y}\ln^{|a_1|-1}\frac{y}{x}\,G_{a_2,\dots,a_\ell}(y)= G_{a_1\wedge a_2,a_3,\dots,a_\ell}(x)\,.
\ee

\section{Proofs of Theorems 1 and 2}
\label{sec:thm}
To prove the Theorems presented in Section \ref{sec:omega} we use of the $x$-space definition of reciprocity equivalent to the  $N$-space relation Eq.~\ref{RRJ}. This is formulated in terms of the Mellin trasform $\tilde{P}(x)$ of $P(N)$
\be\label{parity}
{P}(N)=\int_0^1\, dx\,x^{N-1} \,\tilde{  P}(x)=\M{\tilde{ P}(x)}
\ee
and reads
\be\label{RRmellin}
\widetilde{P}(x)=-x\,\widetilde{P}\left(\textstyle{\frac{1}{x}}\right).
\ee

\subsection{Proof of Theorem 1, no rightmost unit indices}

It is possible to proceed iteratively starting from combinations $\underline{\widehat{\Omega}_a}(N)$ with one index. 
At each step we only focus on $\underline{\widehat{\Omega}}$ combinations with maximal number of indices, the iterative procedure ensures 
in fact that for the remainder the theorem has been already proved.  The strategy is to write the $\underline{\widehat{\Omega}}$ in terms 
of their Mellin transforms exploiting (\ref{Mellinsub}) and use reciprocity in $x$-space via Eq. (\ref{RRmellin}). For this purpose we use 
the notation of Appendix~A and introduce the functions $\Gamma(x)$, whose relation with the $\Omega(N)$ functions is exactly as the one of the 
functions $G(x)$ with the subtracted sums $\underline{\widehat{S}}(N)$. Our derivation mimicks the analogous construction 
described in Sec.~(2.2.1) of~\cite{bdm} generalizing it to the signed case.

\medskip\noindent
For technical reasons, we first consider  $\underline{\widehat{\Omega}_\mathbf{a}}$ in the case where the rightmost index in the 
multi-index $\mathbf{a}$ is not 1. This is necessary since we want to use the Mellin transform described in App.~\ref{subsec:mellin}
which are valid under this limitation. This is not a problem at depth 1 since it is well known that $S_1$ is parity-even. At depth larger than
one, we shall discuss at the end how this limitation can be overcome.  So, let us assume for the moment that 
$\mathbf{a} = (a_1, \dots, a_\ell)$ with $a_\ell \neq 1$.

\medskip\noindent
For one index, 
\be
{\underline{\widehat \Omega_a}}(N)\equiv {\underline{\widehat S_a}}(N)= \vare_{a }^N\,\M{\frac{x}{x-\vare_a}\,G_a(x)}\equiv \vare_{a }^N\, \M{\frac{x}{x-\vare_a}\,\Gamma_a(x)}
\ee
The l.h.s. has parity ${\cal P}=\pm 1$ iff
\be
\Gamma_a(x)={\cal P}\,\vare_a\,\Gamma\left(\textstyle{\frac{1}{x}}\right)
\ee
Using  (\ref{Ggeneral}) it is easy to see that
\be
\vare_a\,\Gamma_a\left(\textstyle{\frac{1}{x}}\right)=(-1)^{|a|-1}\,\vare_a\,\Gamma_a(x)
\ee
Thus, 
 \be\label{cond1}
{\cal P} = (-1)^{|a|-1}\, \vare_a,
 \ee
in agreement with Theorem 1. The generalisation to  $\ell$ indices is straightforward.  
Using the notation $\varepsilon_i \equiv\varepsilon_{a_i}$, it is
\be
{\underline{\widehat \Omega_{a_1,\dots,a_\ell}}}(N)=(\vare_{a_1}\dots\vare_{a_\ell})^N\,
\M{\frac{x}{x-\vare_1\dots\vare_\ell}\Gamma_{a_1,\dots,a_\ell}(x)}
\ee
where
\ba\nonumber
&&\Gamma_{a_1,\dots,a_\ell}(x)= G_{a_1,\dots,a_\ell}(x) 
-{\textstyle\frac{1}{2}}\sum_{k=1}^\ell  G_{a_1,\dots,a_k\wedge a_{k+1}}(x)\\\nonumber
&&+\left(\textstyle{-\frac{1}{2}}\right)^2\Big[
\sum_{k=1}^{\ell-1}G_{a_1,\dots,a_{k-1}\wedge a_k\wedge a_{k+1},\dots,a_\ell}(x)
+\sum_{k=1}^{\ell-2}G_{a_1,\dots,a_{k-1}\wedge a_k,a_{k+1}\wedge a_{k+2},\dots,a_\ell}(x)
\Big]
\\\label{gammaell}
&&+\dots
+\left(\textstyle{-\frac{1}{2}}\right)^{\ell-1}G_{a_1\wedge\dots\wedge a_\ell}(x)\, ,
\ea
which is nothing but the general form of Eq.~(2.17) in~\cite{bdm}.
The l.h.s. has parity ${\cal P}$ iff
\be\label{RR2}
\Gamma_{a_1,\dots,a_\ell}(x)={\cal P} \, \vare_1\dots\vare_\ell \,\Gamma_{a_1,\dots,a_\ell}\left(\textstyle{\frac{1}{x}}\right).
\ee
Using the formula (\ref{Ggeneral}) for each of the $G$-functions evaluated in $1/x$ appearing in the right-hand-side of (\ref{RR2}), one can see that 
\be\label{cond2}
{\cal P} = (-1)^{\sum_{i=1}^\ell(|a_i|-1)}\,\vare_1\dots\vare_\ell,
\ee
again in agreement with Theorem 1 which is then proved for all $\mathbf{a} = (a_1, \dots, a_\ell)$ with $a_\ell \neq 1$.

\subsection{Proof of Theorem 1, extension to general $\mathbf{a}$}

Define the number $u_\mathbf{a}$ of rightmost 1 indices as 
\be
u_\mathbf{a} = \max_k\,\{1\le k \le \ell\,\, |\,\, a_\ell = a_{\ell-1} = \cdots = a_{\ell-k+1} = 1\}.
\ee
One has the identity 
\ba
S_1\,\underline{\widehat\Omega_\mathbf{a}} 
&=& \underline{\Omega_{1, a_1, \dots, a_d}} + \underline{\Omega_{a_1, 1, a_2, \dots, a_d}} + \cdots +  \underline{\Omega_{a_1, \dots, a_d, 1}}+\\
&&   -\frac{1}{4}\,\underline{\Omega_{a_1\wedge a_2 \wedge 1, a_3, \dots, a_d}}
-\frac{1}{4}\,\underline{\Omega_{a_1, a_2 \wedge a_3 \wedge 1, a_4, \dots, a_d}} + \cdots -\frac{1}{4}\,\underline{\Omega_{a_1, \dots, a_{d-2}, a_{d-1} \wedge a_d \wedge 1}}.
\nonumber
\ea
This can be written as
\be
\underline{\Omega_{\mathbf{a},1}} =  S_1\,\underline{\widehat\Omega_\mathbf{a}} + \sum_{\mathbf{b}\in {\cal B}} 
\underline{\Omega_\mathbf{b}},
\ee
where each multi-index $\mathbf{b}\in {\cal B}$ obeys
\be
{\cal P}_\mathbf{b} = {\cal P}_\mathbf{a},\qquad 
u_\mathbf{b} \le u_\mathbf{a}.
\ee
Thus, by induction over $u_\mathbf{a}$ and using the above proof of Theorem 1 for the initial case $u_\mathbf{a}=0$, we 
get the proof of Theorem 1 in the general $u_\mathbf{a}\ge 0$ case.

\subsection{Proof of Theorem 2}

We start from the combinatorial identity
\be\label{omgeneral}
\Omega_{a_1,\dots,a_\ell}(N)=\sum_{k=1}^\ell\,\underline{\widehat{\Omega}_{a_1,\dots,a_k}}(N)\,
\Omega_{a_{k+1},\dots,a_{\ell}}(\infty)+ \Omega_{a_1,\dots,a_\ell}(\infty).
\ee
Suppose now that all even $a_i$ are negative and all odd $a_i$ are positive. Then 
$(-1)^{|a_i|} = -\mbox{sign}(a_i)$ and 
it follows that for any sub-multi-index $(a_1, \dots, a_k)$ we have 
\be
(-1)^{\sum_{i=1}^k(|a_k|-1)}\,\prod_{i=1}^k\,\mbox{sign}(a_i) = (-1)^k\,\prod_{i=1}^k (-1) = 1.
\ee
Thus, from Theorem 1, all terms in the r.h.s. of Eq.~(\ref{omgeneral}) have ${\cal P}=+1$
and Theorem 2 is proved.

\section{Asymptotic expansions of $\gamma$ and $P$}

We report here the first few orders for the large $N$ expansions of the twist-2 anomalous dimension and of its kernel $P$ at four loops.
For the anomalous dimension, using the methods of App.~(\ref{app:asym}), we find 
\ba\nonumber
\gamma_4^{\rm ABA}&=&
-16\Big(\frac{73}{630}\pi^6+4\zeta_3^2\Big) \log \bar N -1400 \zeta_7
-\frac{80}{3} \pi ^2 \zeta_5-\frac{56}{15} \pi ^4 \zeta_3\\\nonumber
&&
+\Big(\frac{96}{5} \pi ^4 \log\bar N+640 \zeta_5-32 \zeta_3^2+\frac{160}{3} \pi ^2 \zeta_3
-\frac{292 \pi^6}{315}\Big)\frac{1}{N}\\\nonumber
&&+\Big(\big(64 \pi ^2-128 \zeta_3 \big)\log ^2\bar N +\big(448
   \zeta_3-\frac{32}{15} \pi ^4 -128 \pi ^2\big) \log
   \bar N\\\nonumber
   &&~~~~-320 \zeta_5+\frac{16 \zeta_3^2}{3}-\frac{32}{3} \pi ^2 \zeta_3
   -384 \zeta_3 +\frac{146 \pi ^6}{945}+\frac{136 \pi ^4}{15}\Big)
   \frac{1}{N^2}\\\nonumber
   &&+\Big(\frac{512  }{3} \log ^3\bar N
   +\big(128\zeta_3-\frac{64}{3} \pi ^2  -768\big) \log ^2\bar N\\\nonumber
   &&~~~~-\big(576 \zeta_3
   +\frac{64}{15} \pi ^4 -\frac{512}{3} \pi ^2 -512\big) \log\bar N \\\label{exp4ABA}
   &&~~~~
   +\frac{320 \zeta_5}{3}-\frac{64}{9} \pi ^2 \zeta_3
   +800 \zeta_3-\frac{32 \pi ^4}{15}-\frac{224 \pi ^2}{3}\Big)
\frac{1}{N^3}+{\cal O}\Big(\frac{1}{N^4}\Big)
\\\nonumber
\gamma_4^{\rm wrapping}&=&-\Big(\frac{64}{3} \pi ^2  +128 \zeta_3  \Big)
\frac{\log^2\bar N}{N^2}  +\Big(\frac{64}{3} \pi ^2  +128 \zeta_3  \Big)\Big(\log^2\bar N-\log \bar N\Big)    \frac{1}{N^3} +{\cal O}\Big (\frac{1}{N^4}\Big)\\\label{exp4wr}&&
\ea
where $\bar N=N\,e^{\gamma_E}$.
The asymptotic next-to-leading constant term is in agreement with~\cite{sbl}, see also~\cite{Fioravanti:2009xt}.
Expanding (\ref{P4ABA}) and (\ref{P4wr}) and summing them together  one obtains the large $N$ expansion of the kernel $P$ at four loops
\ba\label{P4exp}
P_4&=&-16\Big(\frac{73}{630}\pi^6+4\zeta_3^2\Big) \log \bar N -1400
   \zeta_7-\frac{80}{3} \pi ^2 \zeta_5-\frac{56}{15} \pi ^4 \zeta_3-\big(\frac{292 \pi ^6}{315}+32 \zeta_3^2\big)\frac{1}{N}+\\\nonumber
   &&\!\!\!\!\!\!\!\!\!\! -\Big(\big(256 \zeta_3+\frac{64}{3} \pi ^2 \big)\log
   ^2\bar N-\big(64 \zeta_3 +\frac{112}{15} \pi ^4\big) \log\bar N+\frac{8
   \pi ^4}{15}-16 \pi ^2 \zeta_3-\frac{16
   \zeta_3^2}{3}-\frac{146 \pi ^6}{945}\Big) \frac{1}{N^2}+\\\nonumber
   &&\!\!\!\!\!\!\!\!\!\!+\Big(\big(256 \zeta_3+\frac{64}{3} \pi ^2\big) \log ^2 \bar N-\big(320 \zeta_3 +\frac{112}{15} \pi ^4 +\frac{64}{3} \pi ^2\big) \log\bar N-16
   \pi ^2 \zeta_3 +32 \zeta_3 +\frac{64 \pi ^4}{15}\Big)\frac{1}{N^3}
   +{\cal O}\big(\frac{1}{N}\big)^4
   \ea
Notice that  at order $1/N^2$  a $\log^2 N$ appears, which shows the lack of "simplicity" for $P$ and is responsible for the formula (\ref{effc22})  discussed in Section \ref{sec:expansions}.

\end{document}